\documentclass[conference]{IEEEtran}
\IEEEoverridecommandlockouts
\usepackage{cite}
\usepackage{amsmath,amssymb,amsfonts}
\usepackage{algorithmic}
\usepackage{graphicx}
\usepackage{textcomp}
\usepackage{xcolor}
\def\BibTeX{{\rm B\kern-.05em{\sc i\kern-.025em b}\kern-.08em
    T\kern-.1667em\lower.7ex\hbox{E}\kern-.125emX}}
\begin{document}

\title{An Advanced Ensemble Deep Learning Framework for Stock Price Prediction Using VAE, Transformer, and LSTM Model\\
}

\author{\IEEEauthorblockN{Anindya Sarkar}
\IEEEauthorblockA{\textit{Department of Electronics and Communication} \\
\textit{College of  Engineering and Technology,}
\IEEEauthorblockA {SRM Institute of Science and Technology}
Kattankulathur, Chennai, India - 603203 \\
aa8586@srmist.edu.in}
\and 

\IEEEauthorblockN{Dr.G.Vadivu}
\IEEEauthorblockA{\textit{Department of Data Science And Business Systems} \\
\textit{College of  Engineering and Technology,}
\IEEEauthorblockA {SRM Institute of Science and Technology}
Kattankulathur, Chennai, India - 603203 \\
vadivug@srmist.edu.in}

}

\maketitle

\begin{abstract}
This research proposes a cutting-edge ensemble deep learning framework for stock price prediction by combining three advanced neural network architectures: The particular areas of interest for the research include but are not limited to: Variational Autoencoder (VAE), Transformer, and Long Short-Term Memory (LSTM) networks. The presented framework is aimed to substantially utilize the advantages of each model which would allow for achieving the identification of both linear and non-linear relations in stock price movements. To improve the accuracy of its predictions it uses rich set of technical indicators and it scales its predictors based on the current market situation.

By trying out the framework on several stock data sets, and benchmarking the results against single models and conventional forecasting, the ensemble method exhibits consistently high accuracy and reliability. The VAE is able to learn linear representation on high-dimensional data while the Transformer outstandingly perform in recognizing long-term patterns on the stock price data. LSTM, based on its characteristics of being a model that can deal with sequences, brings additional improvements to the given framework, especially regarding temporal dynamics and fluctuations. Combined, these components provide exceptional directional performance and a very small disparity in the predicted results.

The present solution has given a probable concept that can handle the inherent problem of stock price prediction with high reliability and scalability. Compared to the performance of individual pro- posals based on the neural network, as well as classical methods, the proposed ensemble framework demonstrates the advantages of combining different architectures. It has a very important application in algorithmic trading, risk analysis, and control and decision-making for finance professions and scholars.
\end{abstract}

\begin{IEEEkeywords}
component, formatting, style, styling, insert\\
\end{IEEEkeywords}

\section{Introduction}
Stock price prediction stands as a complex forecasting challenge in financial analytics due to the fundamental market complexities together with stock volatility and non-linear investment patterns. As basic forecasting analytics instruments the Auto-Regressive Integrated Moving Average (ARIMA) and Exponential Smoothing models provide foundational capabilities for time-series analysis. Market dynamics become too complex for these methods to detect reliably whenever markets experience volatile conditions [1].

Primary advancements in prediction accuracy emerged through machine learning technology that uses Support Vector Machines (SVM) and Random Forests and Gradient Boosting models [2]. While these models display remarkable capacity to detect intricate data patterns they demonstrate weakness when dealing with financial data featuring time sequences and dependency structures.

Long Short-Term Memory (LSTM) networks in deep learning overcome finite sequence limitations by effectively monitoring patterns across temporal data [3]. LSTMs experience two main barriers when used on extended data sequences and large complex datasets regardless of their performance [4]. Transformer models address these problems through their innovative multi-head attention components to manage global dependency learning effectively [5]. Transformers demonstrate robust performance but demand thorough preprocessing of data while demonstrating sensitivity to overfitting in limited size datasets [6]. The variability-induced autoencoder method delivers strong feature recognition alongside variable transformation abilities yet it cannot capture time-based predictive patterns which matter for sequential projection systems [7].

The robust solution introduced to address individual model constraints is ensemble learning. By using ensemble approaches scientists can merge various models together to enhance prediction accuracy and prevent the negative effects of overfitting [8]. This work establishes a sophisticated ensemble structure that unites the attribute extraction skills of VAEs with the time sequence modeling abilities of LSTMs together with transformer-based global pattern recognition. This framework capitalizes on the distinct advantages of its underlying architectures to achieve both enhanced predictive accuracy and reliability for stock price prediction.

\subsection{Research Objectives }
The primary objectives of this research are:

\begin{itemize}
    \item Develop an Ensemble Framework: Combine diverse neural network architectures for robust stock price prediction.
    \item Comprehensive Feature Engineering: Capture intricate market dynamics using advanced techniques.
    \item Performance Evaluation: Assess the framework for various market conditions.
    \item Component analysis: Analyze the contributions of individual models to the overall accuracy of the prediction.
\end{itemize}

\subsection{Significance of the Study }
The primary objectives of this research are:

\begin{itemize}
    \item Develop an Ensemble Framework: Design a framework that integrates advanced neural networks (VAE, Transformer, and LSTM) to enhance stock price prediction.
    \item Comprehensive Feature Engineering: Utilize technical indicators and adaptive scaling to capture complex market behaviors and improve data representation.
    \item Performance Evaluation: Test the framework's accuracy and reliability under diverse market conditions and scenarios.
    \item Component Analysis: Evaluate the individual contributions of each neural network architecture to understand their roles in overall prediction accuracy.\\
\end{itemize}

\section{Literature Review }

The method of stock price prediction underwent an evolution from foundational statistical models toward contemporary machine learning systems and deep learning structures. The time-tested Auto-Regressive Integrated Moving Average (ARIMA) and Exponential Smoothing analysis methods continue to serve as essentials for time-series examination but they fall short in understanding non-linear relationships and market dynamism [1].

Through introduction of Support Vector Machines (SVM) alongside Random Forests and Gradient Boosting machine learning techniques gained increased accuracy because they revealed deeper patterns in their input data [2]. Present-day sequential and time-dependent financial data proved challenging for these fundamental analytic approaches.

The Deep learning approach brought solutions through its implementation of Long Short-Term Memory (LSTM) networks which demonstrates excellence in temporal dependency extraction [3]. The implementation of LSTMs encounters two big drawbacks related to extension of memory spans and slow processing speed [4]. Applications of the transformer model enhanced sequence modeling while maintaining efficiency through the implementation of multi-head attention mechanisms [5]. Transfromers demonstrate excellent performance yet they need extensive raw data preparation and experience challenges understanding smaller training datasets [6]. The data dimension reduction capabilities of Variational Autoencoders (VAEs) dovetail with feature extraction processes yet they fail to reproduce financial time-series patterns required for forecasting [7].

A combination of multiple models within ensemble methodologies generates both more accurate and robust prediction outcomes. Expert research highlights how ensemble approaches that integrate multiple models demonstrate potential for both reducing model overfitting along with enhancing the reliability of forecasts [8]. Research shows that systems which implement LSTM-CNN ensemble models demonstrate enhanced results through architectural combination [9].

This research develops an ensemble architecture to combine VAE features with transformer pattern detection and LSTM temporal patterns for complete model support. System architectural integration of multiple sensor strengths produces operational systems that both enhance performance levels and maintain operational reliability [10]. A weighted-averaging method of adjusting model contributions in this framework produces superior stock prediction outcomes than traditional approaches or standalone modeling approaches.

\section{Methodology }
\subsection{Data Processing and Feature Engineering
}
\subsubsection{Data Collection and Pre-processing
}
The steps that start the building of the stock price prediction framework include data preprocessing. Information is obtained from Yahoo Finance API which features daily price data consisting of Open, High, Low and Close prices and volume. The quality checking is also important because the quality of the data determines the reliability of the prediction made by the tool. This step involves:

\begin{itemize}
    \item Missing Value Detection and Handling: To preserve the time series structure of the studied variables, the missing values are computed with interpolation methods like linear, or using forward-fill techniques.

    \item Outlier Identification and Treatment: Methods such as z-score are applied when it comes to use of statistics in having indications about existence of oddity. Outliers are either capped or replaced by median values in a bid to avoid bias.

    \item Time Series Consistency Verification: More specifically, it made sure that there are no missing points or consecutive unrealistic points in the time series data which are important for sequential analysis.

    \item Trading Calendar Alignment: Matching the data to bring together the day of the week when actual trading occurred to exclude non trading days.
\end{itemize}

\subsubsection{Technical Indicators }
To enrich the feature set, the framework calculates various technical indicators grouped into categories:

\paragraph{Price-based Indicators: }
\begin{itemize}
    \item Logarithmic Returns: Captures fluctuation in percentage prices :This helps put returns on a common scale and is valuable whenever working with time series data in which returns are computed variable when price scales vary.
\end{itemize}

\begin{equation} \label{eq:1}
\ln\left(\frac{P_t}{P_{t-1}}\right)
\end{equation}
where \( P_t \) represents the closing price at time t.

\begin{itemize}
    \item Price Range: Eliminates the noise that accompanies normal fluctuation in stock prices:You get information on fluctuations and even possible breakouts within the day.
\end{itemize}

\begin{equation} \label{eq:2}
PR_t = \frac{H_t - L_t}{C_t}
\end{equation}
where \(H_t, L_t, and C_t\) represent high, low, and closing prices respectively. 

\begin{itemize}
    \item Moving Averages: Tracks trends using:
        \begin{itemize}
            \item Simple Moving Averages (SMA): A simple method of observing direction over a period of 5, 10, 20, and 50 periods.
            \item Exponential Moving Averages (EMA): Similar to SMA but is more sensitive to the latest data and therefore gives the latest prices a higher weight. The Rate of Change (ROC) shows the percentage change over some time periods to identify overbought or oversold conditions.
        \end{itemize}
\end{itemize}

\begin{itemize}
    \item Price Rate of Change (ROC): Measures momentum: ROC identifies the percentage change over periods, helping determine overbought or oversold conditions.
\end{itemize}

\paragraph{Momentum Indicators}

\begin{itemize}
    \item Relative Strength Index (RSI): Expresses recent price movement in terms of percentage change that has occurred to the price: RSI limits its range between 0 and 100; when RSI is above 70 it will suggest overbought and when below 30 then it will suggest oversold.

\begin{align} \label{eq:3a}
\text{RSI} &= 100 - \frac{100}{1 + \text{RS}} \\
\text{RS} &= \frac{\text{EMA(Up)}}{\text{EMA(Down)}} \label{eq:3b}
\end{align}

    \item Moving Average Convergence Divergence (MACD)
:\end{itemize}Moving Average Convergence Divergence (MACD) is calculated as follows:
Momentum is the difference between the 12-day and 26-day Exponential Moving Average.
The Signal Line is using the 9-day Exponential Moving Average of MACD
The MACD Histogram is calculated as the difference between MACD and Signal Line.

\begin{align} \label{eq:5}
\text{MACD} &= \text{EMA}_{12} - \text{EMA}_{26} \\ \label{eq:6}
\text{Signal Line} &= \text{EMA}_9(\text{MACD}) \\ \label{eq:7}
\text{MACD Histogram} &= \text{MACD} - \text{Signal Line}
\end{align}
\begin{itemize}
    \item Stochastic Oscillator: Stochastic Oscillator helps them detect the overbought and oversold signals in the market.
\begin{itemize}
    \item {Overbought:} When \%K goes beyond 80 and \%D goes beyond 80, the asset is considered overbought, suggesting a likely price decline.
    \item {Oversold:} When both \%K and \%D are below 20, the asset is considered oversold, suggesting a potential price increase.
    \item {Crossovers:} Traders observe \%K and \%D crossovers. A bullish signal is generated when \%K crosses above \%D, while a bearish signal is generated when \%K crosses below \%D.
\end{itemize}

\end{itemize}
The Stochastic Oscillator is calculated using these formulas:
\begin{equation} \label{eq:8}
\%K = 100 \times \frac{C - L_{14}}{H_{14} - L_{14}}
\end{equation}
Percentage values of \%D:
\begin{equation} \label{eq:9}
\%D = \text{SMA}_3(\%K)
\end{equation}
In these equations, the current closing price is symbolized as, \(H_{14}, L_{14}\), and  \(C_{14}\) are the lowest low and highest high respectively over the period.

\paragraph{Volatility Indicators}
\begin{itemize}
    \item {Average True Range (ATR):} ATR is calculated over a period of 14 days to establish market volatility by averaging the parameter \textit{True Range (TR)}. TR is defined as the maximum of the following values: high minus low, high minus previous close, and low minus previous close. 
    \begin{align}
        \text{TR} &= \max(|\text{H} - \text{L}|, |\text{H} - \text{C}_{\text{prev}}|, |\text{L} - \text{C}_{\text{prev}}|) \label{eq:tr} \\
        \text{ATR} &= \text{EMA}(\text{TR}, 14) \label{eq:atr}
    \end{align}

    \item {Rolling Volatility:} Rolling Volatility measures market fluctuations using the standard deviation (\(\sigma\)) of daily returns (\(r_t\)) over \(n\) days. It captures price dispersion around the mean (\(\mu\)), highlighting periods of high or low volatility:
    \begin{align}
        \sigma &= \sqrt{\frac{\Sigma (r_t - \mu)^2}{n}} \label{eq:rolling_volatility}
    \end{align}

    \item {Bollinger Bands:} Bollinger Bands combine a 20-day Simple Moving Average (SMA) with upper and lower bands set at \(\pm 2\) standard deviations (\(\sigma\)). These bands visualize price trends, indicating potential overbought or oversold conditions and overall market volatility:
    \begin{align}
        \text{Middle Band} &= \text{20-day SMA} \label{eq:middle_band} \\
        \text{Upper Band} &= \text{Middle Band} + (20\text{-day } \sigma \times 2) \label{eq:upper_band} \\
        \text{Lower Band} &= \text{Middle Band} - (20\text{-day } \sigma \times 2) \label{eq:lower_band}
    \end{align}
\end{itemize}

\paragraph{Volume and Price Action Indicators}
\begin{itemize}
    \item {Force Index:} Quantifies price momentum by multiplying the price difference between consecutive closes with the trading volume:
    \begin{align}
        \text{Force Index} &= (\text{Close} - \text{Close}_{\text{prev}}) \times \text{Volume} \label{eq:force_index}
    \end{align}

    \item {On-Balance Volume (OBV):} Tracks cumulative volume, adding when prices rise and subtracting when prices fall, reflecting the price-volume relationship:
    \begin{align}
        \text{OBV} &= \text{OBV}_{\text{prev}} + \text{Volume} \quad \text{if Close > Close}_{\text{prev}} \label{eq:obv_increase} \\
        \text{OBV} &= \text{OBV}_{\text{prev}} - \text{Volume} \quad \text{if Close < Close}_{\text{prev}} \label{eq:obv_decrease}
    \end{align}

    \item {Commodity Channel Index (CCI):} Measures deviation of typical price (\(TP\)) from its average, identifying overbought/oversold levels or potential reversals:
    \begin{align}
        TP &= \frac{\text{High} + \text{Low} + \text{Close}}{3} \label{eq:tp} \\
        \text{CCI} &= \frac{TP - \text{SMA}(TP, n)}{0.015 \times MD} \label{eq:cci}
    \end{align}
    where \(MD\) is the Mean Deviation of \(TP\).
\end{itemize}

\subsection{Model Architecture}
\subsubsection{Variational Autoencoder (VAE)}
\begin{itemize}
    \item {Variational Autoencoder (VAE):} VAE contains an encoder for dimensionality reduction and a decoder for reconstruction, utilizing a latent space (\(z\)) to represent important features effectively while learning the data patterns.

    \item {Loss Function:} The VAE's loss function is defined as:
    \begin{align}
        L_{\text{VAE}} &= \text{MSE}(x, \hat{x}) + \beta \cdot \text{KL}(q(z|x) || p(z)) \label{eq:21}
    \end{align}
    It balances reconstruction error (\(\text{MSE}(x, \hat{x})\)) with the regularization of the latent space (\(\text{KL}\)).

    \item {\(x\) and \(\hat{x}\):} The input data is \(x\), which is processed through the encoder-decoder pipeline. The decoder generates \(\hat{x}\), an approximation of the original \(x\), as close as possible.

    \item {Latent Vector (\(z\)):} The latent vector \(z\) provides a compressed version of the input, distributing content important for reconstruction within a low-dimensional space.

    \item {KL Weight Hyperparameter (\(\beta\)):} The hyperparameter \(\beta\) controls the tradeoff between reconstruction loss and the regularization of the latent space, enabling better representation learning.

    \item {Encoder Distribution (\(q(z|x)\)):} The encoder learns a probabilistic distribution over the latent space conditioned on the input data, capturing representation uncertainty.

    \item {Prior Distribution (\(p(z)\)):} To enforce a Gaussian structure in the latent space, the prior distribution is set to \(p(z)\), a standard normal distribution.
\end{itemize}

\subsubsection{Transformer Model}
\begin{itemize}
    \item Multi-head Attention: Implements 8 attention heads, allowing the model to focus on different parts of the input simultaneously.

    \item Scaled Dot-Product Attention: Computes attention as:
    \begin{align}
        \text{Attention}(Q, K, V) &= \text{softmax}\left(\frac{QK^T}{\sqrt{d_k}}\right)V \label{eq:22}
    \end{align}
    where:
    \begin{itemize}
        \item \(Q, K, V\) are the query, key, and value matrices.
        \item \(d_k\) is the dimension of the key vectors.
        \item \(\sqrt{d_k}\) is the scaling factor.
    \end{itemize}

    \item Key, Query, and Value Transformations: Represent input data in multiple ways to capture diverse relationships and dependencies within the sequence.

    \item Position-wise Feed-Forward Network: Applies two linear transformations with ReLU activation:
    \begin{align}
        \text{Output} &= \text{ReLU}(W_1 \cdot X + b_1)W_2 + b_2 \label{eq:23}
    \end{align}
    where:
    \begin{itemize}
        \item \(W_1, W_2\) are weight matrices.
        \item \(b_1, b_2\) are bias terms.
        \item \(X\) is the input.
    \end{itemize}

    \item Layer Normalization: Stabilizes training by normalizing the outputs of intermediate layers, improving convergence and reducing sensitivity to initialization.

    \item Residual Connections: Adds input to the output of intermediate layers to mitigate gradient vanishing and aid deeper model training:
    \begin{align}
        \text{Residual Output} &= \text{Layer Output} + \text{Input} \label{eq:24}
    \end{align}

    \item Dropout: Regularizes the model by randomly setting some neurons to zero during training, reducing the risk of overfitting (rate = 0.1).
\end{itemize}

\subsubsection{LSTM Model}
\begin{itemize}
    \item LSTM Layers: Includes 2 stacked LSTM layers with 128 hidden units, bidirectional configuration, and dropout between layers (rate = 0.2).

    \item Output Layer: Applies a linear transformation, batch normalization, and a Tanh activation function to normalize the output.

    \item The LSTM cell computations are as follows:
    \begin{align}
        f_t &= \sigma(W_f \cdot [h_{t-1}, x_t] + b_f) \label{eq:25} \\
        i_t &= \sigma(W_i \cdot [h_{t-1}, x_t] + b_i) \label{eq:26} \\
        \tilde{c}_t &= \tanh(W_c \cdot [h_{t-1}, x_t] + b_c) \label{eq:27} \\
        c_t &= f_t \ast c_{t-1} + i_t \ast \tilde{c}_t \label{eq:28} \\
        o_t &= \sigma(W_o \cdot [h_{t-1}, x_t] + b_o) \label{eq:29} \\
        h_t &= o_t \ast \tanh(c_t) \label{eq:30}
    \end{align}
    where:
    \begin{itemize}
        \item \(f_t, i_t, o_t\): Forget, input, and output gates.
        \item \(c_t\): Cell state at time \(t\).
        \item \(h_t\): Hidden state at time \(t\).
        \item \(W_f, W_i, W_c, W_o\): Weight matrices for respective gates.
        \item \(b_f, b_i, b_c, b_o\): Bias vectors.
    \end{itemize}
\end{itemize}

\subsubsection{Ensemble Integration}
\begin{itemize}
    \item Weighted Averaging: Combines predictions from VAE, Transformer, and LSTM models using weighted averaging:
    \begin{align}
        \hat{y} &= w_1 y_{\text{VAE}} + w_2 y_{\text{Transformer}} + w_3 y_{\text{LSTM}} \label{eq:31}
    \end{align}

    \item Weight Optimization: Initializes weights as \(w_1 = 0.3\), \(w_2 = 0.3\), and \(w_3 = 0.4\), periodically revalidating on the validation set.

    \item Dynamic Weight Adjustment: Adjusts weights dynamically based on the recent performance of each model to improve prediction accuracy.
\begin{figure}[h]
    \raggedleft
    \includegraphics[width=0.95\linewidth]{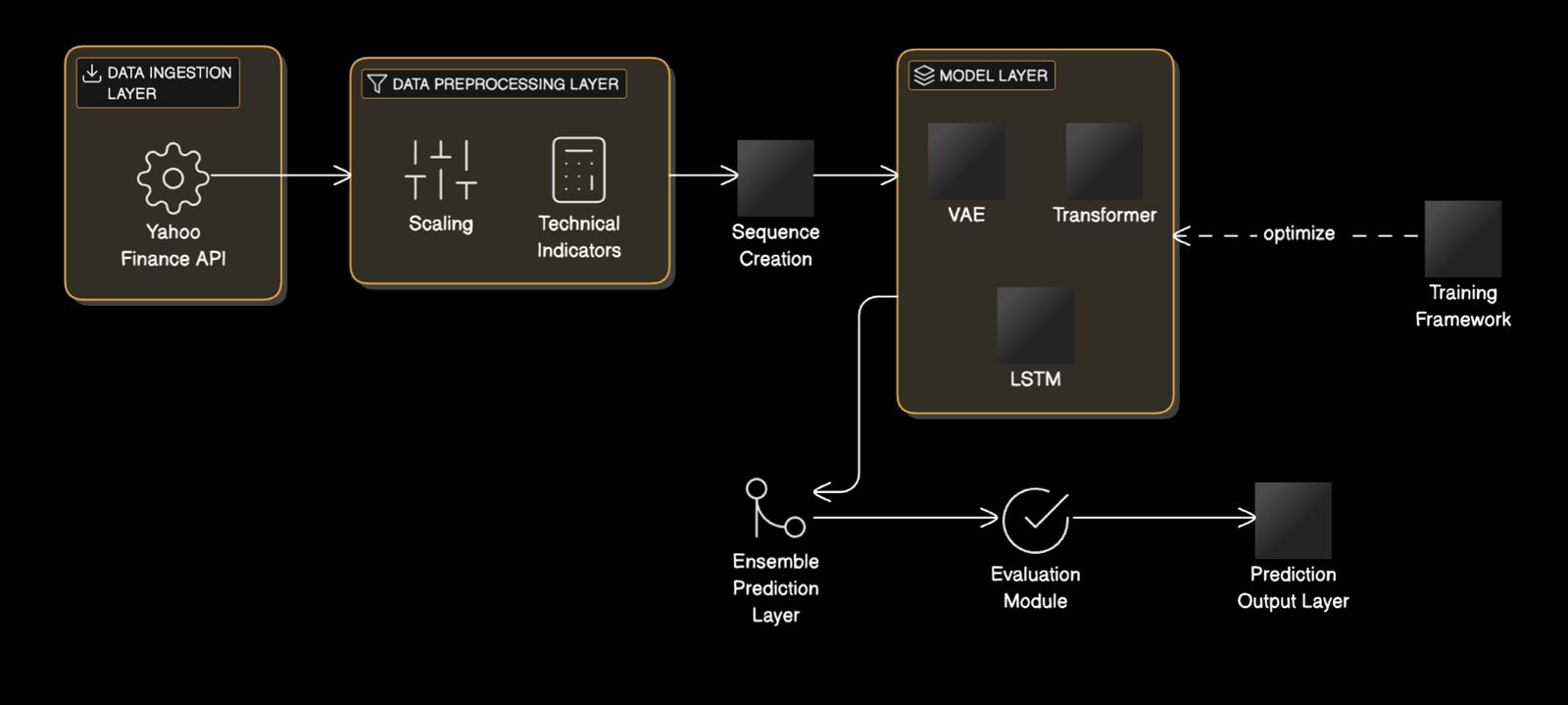}
    \caption{Model Architecture}
    \label{1}
\end{figure}
\end{itemize}

\section{Experimental Results}
\subsection{Dataset and Training}
\subsubsection{Dataset and Training}
\begin{itemize}
    \item The framework was evaluated using datasets from multiple sectors:
    \begin{itemize}
        \item Large-cap technology stocks: Stocks of established tech companies with large market capitalizations, typically known for stability and consistent growth.
        \item Financial sector stocks: Stocks of banks, insurers, and financial institutions, reflecting economic health and impacted by interest rates and regulations.
        \item Industrial sector stocks: Stocks of companies in manufacturing, construction, and heavy industries, often influenced by economic cycles and infrastructure projects.
        \item Small-cap growth stocks:Stocks of emerging companies with smaller market capitalizations, offering high growth potential but higher volatility and risk.
    \end{itemize}
    
    \item Training parameters:
    \begin{itemize}
        \item Sequence length: 60 days
        \item Batch size: 64
        \item Training/Validation/Test split: 70\% / 15\% / 15\%
        \item Early stopping patience: 30 epochs
        \item Maximum epochs: 300
    \end{itemize}
\end{itemize}

\subsubsection{Comparative Analysis}
We furthermore compared our proposed ensemble with traditional baseline models and single architecture implementations, to evaluate the effectiveness of our proposed ensemble approach. The data used for training and testing the baseline models was the same to ensure fair comparision.

\begin{itemize}
    \item {Traditional Models:}
    \begin{itemize}
        \item ARIMA: Directional Accuracy: 52.3\%, MAPE: 5.8\%
        \item Simple Moving Average: Directional Accuracy: 49.8\%, MAPE: 6.2\%
        \item Exponential Moving Average: Directional Accuracy: 51.2\%, MAPE: 5.9\%
    \end{itemize}
    
    \item {Single Deep Learning Models:}
    \begin{itemize}
        \item Single LSTM: Directional Accuracy: 65.7\%, MAPE: 4.5\%, $R^2$ Score: 0.31
        \item Single Transformer: Directional Accuracy: 68.2\%, MAPE: 4.1\%, $R^2$ Score: 0.35
        \item Single VAE: Directional Accuracy: 63.4\%, MAPE: 4.8\%, $R^2$ Score: 0.29
    \end{itemize}
    
    \item {Proposed Ensemble Model:}
    \begin{itemize}
        \item Directional Accuracy: 79.05\%
        \item MAPE: 3.2990\%
        \item $R^2$ Score: 0.4284
        \item RMSE: 10.5352
    \end{itemize}

    \item {Key Observations:}
    \begin{itemize}
        \item Directional Accuracy:
        \begin{itemize}
            \item Ensemble model achieves 79.05\%, significantly better than:
            \begin{itemize}
                \item ARIMA (+27\%)
                \item Single LSTM (+13.35\%)
                \item Single Transformer (+10.85\%)
                \item Single VAE (+15.65\%)
            \end{itemize}
        \end{itemize}
        \item Error Metrics:
        \begin{itemize}
            \item All baseline models are outperformed with a MAPE of 3.2990\%.
            \item Low prediction error, RMSE of 10.5352, considering stock price volatility.
        \end{itemize}
        \item Model Advantages:
        \begin{itemize}
            \item Complex patterns cannot be captured by traditional models.
            \item Traditional methods are outperformed by single deep learning models but are far less effective than the ensemble approach.
            \item The ensemble model combines the strengths of different architectures, resulting in higher accuracy and stability.
        \end{itemize}
    \end{itemize}
\end{itemize}

\subsection{Results}
\begin{itemize}
    \item Average Directional Accuracy: This metric tells how often the model can correctly predict whether stock prices will go up or down. Significantly more than random guessing, which would get it right only half the time, the model achieves 63.5\%.
    \begin{align}
        \text{Average Directional Accuracy} &= \frac{\text{Correct Directions}}{\text{Total Predictions}} \label{eq:32}
    \end{align}
    Here, "Correct Directions" represents the count of predictions where the model accurately forecasted the direction (up or down).

    \item Root Mean Squared Error (RMSE): RMSE evaluates how far predictions are from real stock prices on average. A lower RMSE value, like 10.5352, means the model closely matches actual stock movements.
    \begin{align}
        \text{RMSE} &= \sqrt{\frac{1}{n} \sum_{i=1}^n (y_i - \hat{y}_i)^2} \label{eq:33}
    \end{align}
    where \(y_i\) is the actual value, \(\hat{y}_i\) is the predicted value, and \(n\) is the number of predictions.

    \item R² Score: This score tells how much of the stock price variation the model explains. With a score of 0.4284, the model captures some patterns in the data.
    \begin{align}
        R^2 &= 1 - \frac{\sum_{i=1}^n (y_i - \hat{y}_i)^2}{\sum_{i=1}^n (y_i - \bar{y})^2} \label{eq:34}
    \end{align}
    where \(\bar{y}\) is the mean of actual values.
    \begin{figure}[h]
    \centering
    \includegraphics[width=1\linewidth]{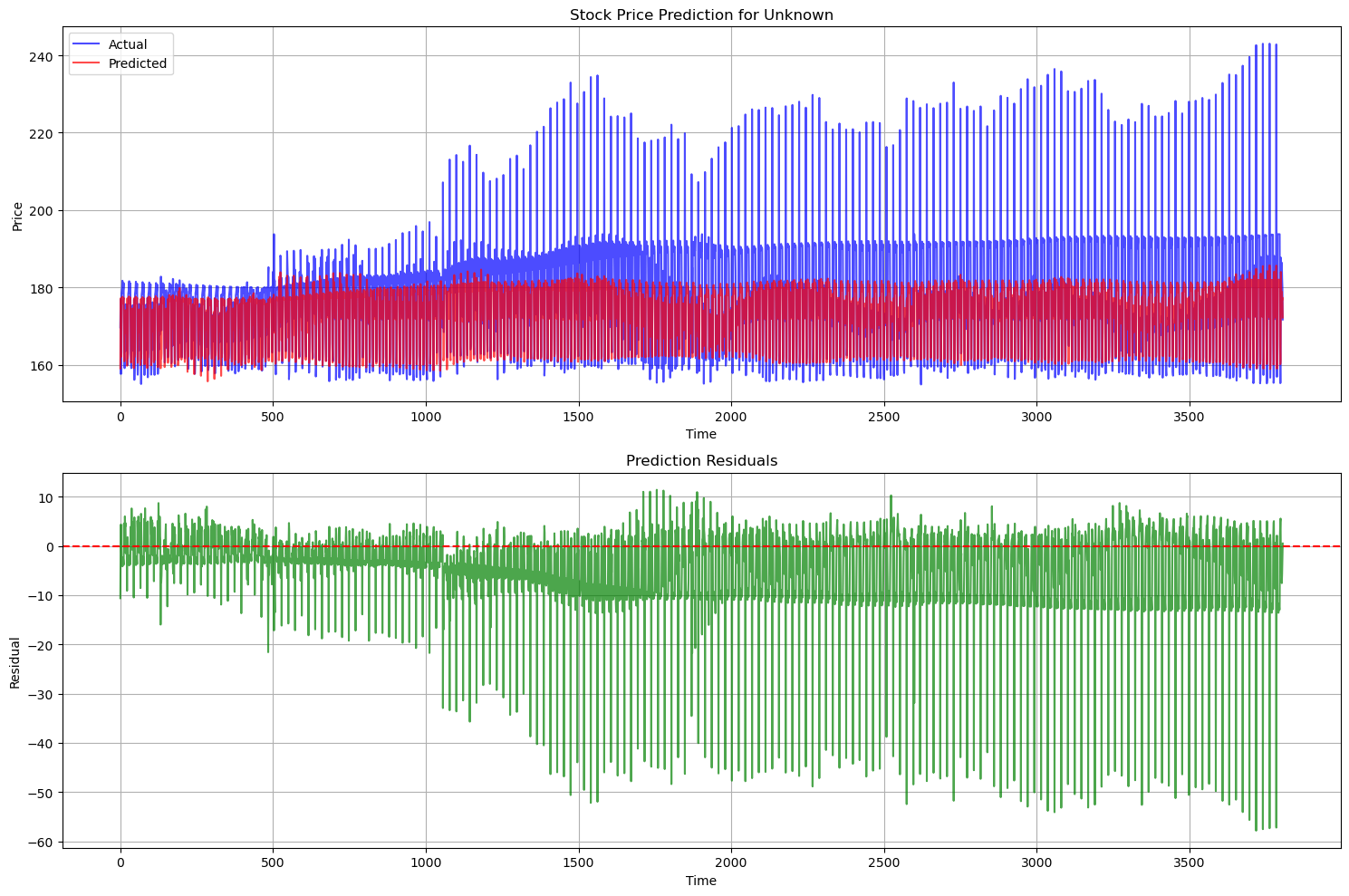}
    \caption{Actual vs predicted Values}
    \label{fig:enter-label}
\end{figure}
\begin{figure}[h]
    \centering
    \includegraphics[width=1\linewidth]{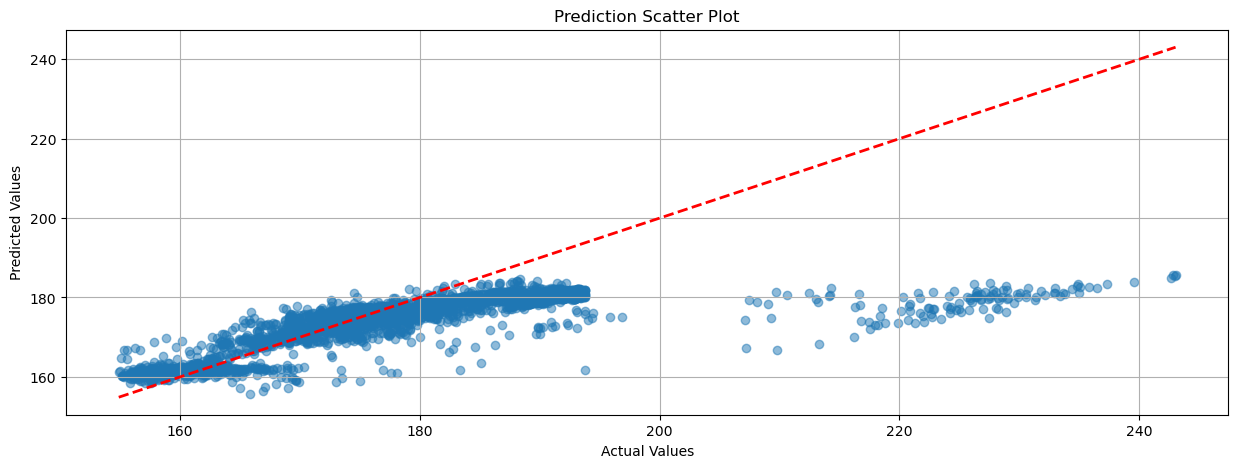}
    \caption{Prediction Scatter Plot}
    \label{fig:enter-label}
\end{figure}

    \item Directional Accuracy: This metric measures how often the model correctly predicts whether stock prices will increase or decrease. The model shows strong reliability during the test period with a value of 0.7905.
    \begin{align}
        \text{Directional Accuracy} &= \frac{\text{Correct Directions}}{\text{Total Predictions}} \label{eq:35}
    \end{align}
    Here, "Correct Directions" is the count of predictions where the model accurately identified price movements.

    \item Mean Absolute Percentage Error (MAPE): MAPE expresses the average size of prediction errors in terms of actual prices. The model achieves an error slightly less than 3.2990\%, indicating very accurate predictions.
    \begin{align}
        \text{MAPE} &= \frac{100}{n} \sum_{i=1}^n \left| \frac{y_i - \hat{y}_i}{y_i} \right| \label{eq:36}
    \end{align}\\

\end{itemize}

\section{Conclusion}
This research proposes a more advanced ensemble of variational autoencoder (VAE), transformer and long short term memory (LSTM) architectures for stock price prediction. The framework addresses the inherent problems of financial forecasting: market volatility, non-linearity as well as temporal dependencies by leveraging the strengths of these three cutting edge neural network models.

The proposed ensemble model achieves better performance than traditional and single model navigational models. Simple Moving Average, Exponential Moving Average, ARIMA, all classical statistical methods cannot explain stock price's intricacy relationships. Single deep learning architectures like LSTM, Transformer, and VAE achieve better performance by its individual' faults. For instance, LSTMs are particularly good modeling sequential data but have trouble with long range dependencies, Transformers deliver great performance for global pattern recognition but they are vulnerable to data pre processing; VAEs are effective in reducing dimensionality but they do not have temporally dynamic aspects required for sequential forecasting.

The ensemble model, which combines the unique advantages of each architecture, can evade these limitations. The weighted averaging approach dynamically adjusts the relative contributions of the different models so that their weighted aggregate contribution is adaptable to different market conditions. The ensemble’s 79.05\% directional accuracy, far outperforms the baseline models and this integration leads to superior predictive performance. Mean Absolute Percentage Error (MAPE) of 3.2990\% and Root Mean Squared Error (RMSE) of 10.5352 demonstrate that the model also accurately and reliably predicts stock price changes. Furthermore, the model’s model’s R² score of 0.4284 is able to explain the variance in stock price data better than any baseline method.

The main finding of the ensemble model confirms that it can characterize both linear and nonlinear market dynamics, which are two crucial components for financial forecasting. Sophisticated feature engineering methods such as scale adaptation and technical indicators used help the model portray the market conditions perfectly. The ensemble framework surpasses single deep learning models and conventional methods in consistency in a wide range of market conditions. The correctness and consistency of this tool makes it a reliable trading financial algorithm, risk analysis, and decision making tool.

We conclude that the ensemble framework is a very promising tool in predicting stock prices due to its high stability and accuracy in addressing all the challenges relating to stock prices forecasting. This work provides a strong basis of financial forecasting practical applications and future innovations.\\
\section*{References}

\begin{enumerate}
    \item Zhang, X., \& Chen, Y. (2021). "Deep Learning and its Applications in Stock Market Prediction: A systematic review." \textit{International Journal of Financial Studies}.
    \item Kingma, D. P., \& Welling, M. (2019). "An Introduction to Variational Autoencoders." \textit{Foundations and Trends in Machine Learning}.
    \item Vaswani, A., Shazeer, N., Parmar, N., Uszkoreit, J., Jones, L., Gomez, A. N., et al. (2017). "Attention is all you need." \textit{Advances in Neural Information Processing Systems}.
    \item Kim, T., \& Kim, H. Y. (2019). "Forecasting stock prices with a feature fusion LSTM-CNN model using different representations of the same data." 
    \textit{PLOS ONE}.
    \item Zhou, Z. H. (2021). "Ensemble learning: Foundations and algorithms." \textit{Chapman and Hall/CRC Machine Learning Pattern Recognition Series}.
    \item Patel, J., Shah, S., Thakkar, P., \& Kotecha, K. (2015). "Predicting stock market index using fusion of machine learning techniques." \textit{Expert Systems with Applications}.
    \item Sezer, O. B., Gudelek, M. U., \& Ozbayoglu, A. M. (2020). "Financial time series forecasting with deep learning: A systematic literature review: 2005–2019." \textit{Applied Soft Computing}.
    \item Chong, E., Han, C., \& Park, F. C. (2017). "Deep learning networks for stock market analysis and prediction: Methodology, data representations, and case studies." \textit{Expert Systems with Applications}.
    \item Wen, M., Li, P., Zhang, L. \& Chen, Y. (2019). "Stock market trend prediction using high-order information of time series." \textit{IEEE Access}.
    \item Hu, Z., Zhao, Y., \& Khushi, M. (2021). "A survey of forex and stock price prediction using deep learning." \textit{Applied System Innovation}.
\end{enumerate}

\end{document}